\documentstyle[aps,prl,multicol,epsf]{revtex}

\def\R{\hbox{\rm I \kern-5pt R}}

\title{A proposal for founding mistrustful quantum cryptography on
  coin tossing}

\author{Adrian Kent} 
\address{ 
Centre for Quantum Computation, DAMTP, Centre for Mathematical Sciences,\\
University of Cambridge, Cambridge CB3 0WA, U.K.${}^{1}$ \\
{\rm and}\\
Hewlett-Packard Laboratories, Filton Road,\\
Stoke Gifford, Bristol BS34 8QZ, U.K.
}
\date{March 2003 (revised)} 

\begin{document} 
\maketitle
\begin{abstract}
A significant branch of classical cryptography deals with the problems
which arise when mistrustful parties need to generate, 
process or exchange information.  As Kilian showed a while
ago, mistrustful classical cryptography can be founded on 
a single protocol, oblivious transfer, from which general
secure multi-party computations can be built.  

The scope of mistrustful quantum cryptography is limited by no-go
theorems, which rule out, inter alia, unconditionally secure quantum
protocols for oblivious transfer or general secure two-party
computations.  These theorems apply even to protocols which take
relativistic signalling constraints into account.  The best that can
be hoped for, in general, are quantum protocols computationally secure
against quantum attack.  I describe here a method for building a
classically certified bit commitment, and hence every other
mistrustful cryptographic task, from a secure coin tossing protocol.
No security proof is attempted, but I sketch reasons why these
protocols might resist quantum computational attack.
\end{abstract}
\vskip 5pt
${}^1$ Present and permanent address

\begin{multicols}{2}

\section{Introduction} 

Quantum computers pose a threat to most, if not all, standard 
classical cryptographic schemes.  Typically, classical cryptosystems
rely on the difficulty of factorisation or equivalent tasks, which
we know quantum computers can solve efficiently.  Even classical 
protocols which rely on problems for which no efficient quantum 
algorithm is currently known are somewhat suspect at the moment, 
since the power of quantum computers is not well understood.   

Fortunately, for key distribution and a few other interesting
cryptographic tasks, quantum cryptography offers a complete defence
to the threat posed by quantum computers --- unconditionally secure
quantum protocols, which are provably unbreakable by classical or 
quantum computers.  Unfortunately, this is not true of a wide range 
of important cryptographic tasks that allow mistrustful parties 
to generate, process or exchange information with suitable 
security guarantees.   No-go theorems show the impossibility
of unconditionally secure non-relativistic quantum protocols 
for many of these tasks --- for example, bit 
commitment\cite{mayersprl,mayerstrouble,mayersone,lochauprl,lochau}, oblivious 
transfer and some secure two-party computations\cite{lo}.  In the last
two cases, these theorems apply also to protocols which take
account of relativistic signalling constraints.   

Given that unconditional security is unattainable for these tasks, 
we have to fall back on weaker notions of security. 
One possible approach is to assume that that reliable 
bounds can be placed on the size of any quantum computer in the
possession of an adversary, and to devise protocols which 
cannot be broken by quantum computers capable of manipulating
no more than $N$ qubits coherently\cite{salvail}. 
However, it is hard to tell at the moment whether future
technological developments will allow for any such bounds. 
It is also known that bit commitment protocols can be devised
which are secure under the assumption that quantum one-way
functions exist.\cite{dms}  However, identifying good candidate quantum
one-way functions is itself a challenge.  

It would, at any rate, certainly be good to be able to return to the 
``pre-quantum'' state of affairs, replacing protocols which offer
computational security against classical computers with protocols
which offer credible computational security against quantum
computers.   For example, since NP-complete problems are generally thought
unlikely to be solvable by quantum computers in polynomial
time one might hope to build protocols 
whose security relies on the difficulty of solving a particular
instance of a problem whose general case is NP-complete.    
(An NP-complete problem has the property that every other problem
in NP can be polynomially reduced to it: in other words it is at
least as hard as any problem in NP.) 

But there is an obvious difficulty here.  Mistrustful
cryptographic problems require security against both parties.   
But if $A$ proposes using a particular instance of a 
problem, $B$ has no way of verifying for sure that the 
particular problem proposed genuinely {\it is} hard. 
It is presumably in $A$'s interests to choose, if she can, an 
apparently hard problem with hidden structure, which is itself very
hard to find, but which allows $A$ to solve the problem easily.   

In this letter, I propose and briefly discuss a method 
which offers a possible way round this obstacle: 
using remote coin tossing --- which we know can be implemented with
perfect security by using relativistic signalling constraints,
and with good computational security without making use of
relativity --- to allow mistrustful parties to generate random instances
of hard problems.   

I focus on one particularly important protocol --- classically
certified bit commitment.  In a classically certified bit commitment
protocol, $B$ is guaranteed that $A$ is committed to some fixed
classical bit value, $0$ or $1$: the possibility that $A$'s bit
commitment is described, until unveiling, by a quantum mixture
of $0$ and $1$ can be excluded.  
Classically certified bit commitment is stronger than ordinary
bit commitment in the quantum realm, and cannot be implemented
with unconditional security even by quantum protocols which take account of
relativistic signalling constraints\cite{kentbccc}. 

It is known\cite{yao,kilian}
that secure oblivious transfer and, hence, general secure multi-party
computation can be implemented by quantum information exchanges,
given a secure classically certified bit commitment protocol.   

\section{Remote coin tossing}

A remote coin tossing protocol was originally defined\cite{blum} as a 
protocol that allows two
mistrustful parties, A and B, to generate a random bit, in such a way
that each has confidence that, so long as they behave honestly (whether
or not the other party did), the resulting bit $b$ is genuinely random.
In fact, as Mayers et al. pointed out\cite{msc}, 
this seemingly straightforward definition hides a subtlety.
A coin tossing protocol could fail to generate any bit, if one of 
the parties chooses to abort the protocol before it is complete.
A cheating party might preferentially tend to abort if the output
bit appears likely to take the value disfavoured by that party.  

To allow for this possibility, we follow Mayers et al.\cite{msc} in defining
a secure coin tossing protocol to be one which guarantees the following.
First, if both parties are honest, then ${\rm Prob} ( b = 0 ) = 
{\rm Prob} (b=1 ) = 1/2$.  Second, if one party is honest, then, 
whatever strategy the dishonest party uses, $ {\rm Prob} ( b = 0 ) < 1/2 +
\epsilon$ and $ {\rm Prob} ( b = 1 ) < 1/2 + \epsilon$.  
An {\it ideal} coin tossing protocol guarantees this with $\epsilon = 0$.
A secure coin tossing protocol need not be ideal, so long as it contains
parameters which can be chosen so as to make $\epsilon$ as small as
desired.   

Coin tossing is known to be strictly weaker than bit commitment 
in non-relativistic classical and quantum cryptography.
The reason is that any secure bit 
commitment protocol can be used for secure coin tossing:
$A$ commits a random bit $a$ to B; B returns a random bit $b$; 
$A$ then unveils $a$ and they take $a \oplus b$ as the coin toss
outcome.  On the other hand, it is impossible to build a secure 
non-relativistic classical or quantum bit commitment protocol using a
black box for secure ideal coin tossing.\cite{akweaker}

A simple unconditionally secure ideal coin tossing protocol can be defined by
using relativistic signalling constraints.   Fix some inertial
coordinates, agreed by $A$ and $B$.  Suppose $A$ controls sites
$A_1$ and $A_2$, and $B$ controls sites $B_1$ and $B_2$, such that
$A_1$ and $B_1$ are within distance $\delta$ of some agreed 
point $P_1$ and $A_2$ and $B_2$ are within distance $\delta$ of some agreed 
point $P_2$, where $P_1$ and $P_2$ are separated by $d \gg \delta$.  
Fix also some time $t$ agreed by $A$ and $B$.  At time $t$, $A_1$ sends
a random bit $a$ as a classical signal, to be received by $B_1$; at
the same time, $B_2$ sends a random bit $b$ as a classical signal,
to be received by $A_2$.   $A$ and $B$ accept these signals as 
valid implementations of the protocol provided they are received
(by $B_1$ and $A_2$ respectively), by time $t + 2 \delta$.   
They then take $a \oplus b$ as the coin toss outcome.   
Each party is guaranteed that the outcome
is randomly generated, so long as they receive the other party's
bit at a point outside the future light cone of the point from
which their own was transmitted, and regardless of whether the 
other party's chosen bit was genuinely random.   

Relativistic coin tossing at a high bit rate is eminently practical,
but will not work if the parties are unable or unwilling to arrange 
to control suitably adjacent separated sites.  
Strangers communicating by phone or over the internet, for instance, 
are likely to need a non-relativistic protocol if they urgently
need to generate random bits.  
Unfortunately, no unconditionally secure non-relativistic classical 
coin tossing protocol exists.  It is claimed\cite{kitaev} that one can
demonstrate that no unconditionally secure non-relativistic quantum coin tossing protocols
exist: to the best of my knowledge, no written proof has as yet been 
circulated.  It is known that unconditionally secure ideal quantum
coin tossing is impossible.\cite{lochau}

Another approach to secure coin tossing is to build a coin tossing
protocol from a bit commitment
protocol which $A$ and $B$ trust to be temporarily computationally secure.  
They can then use these temporarily secure bit commitments to
implement computationally secure coin tossings, using the 
construction described above --- so long as the bit commitment
is trusted to be secure for as long as it takes to exchange 
messages.    

Even in a future world where large quantum computers
are commonplace, it might be reasonable to
have great confidence in the temporary security --- for, say, a
few seconds --- of standard classical bit commitments.
Essentially, this requires problems which, one can be confident,
take considerably longer to solve than to state and communicate.   
Problems which are only polynomially hard for quantum
computers, such as factorisation, might well suffice.  

If temporarily secure bit commitments are used, the protocols
below effectively define a form of bootstrapping, in which bit 
commitments that are (plausibly) computationally secure 
for a very long time are built from secure coin tossings,
which themselves are built on bit commitments that are 
computationally secure only for a relatively short time.  
 
In any case, in the rest of this paper it is assumed that some trusted secure 
remote coin tossing method is available to $A$ and $B$.  This
could be the relativistic scheme described above, a scheme
that is trusted to be computationally secure, or an
(as yet undiscovered) unconditionally secure quantum coin
tossing scheme that does not rely on relativistic signalling
constraints --- or any other scheme whose security can be
trusted.  Whichever, we assume that the security of the
scheme extends to multiple coin tosses, in the sense that 
the participants can trust that implementing the scheme $N$ 
times is approximately
equivalent to sampling $N$ independent and identically
distributed random variables, each corresponding to a fair coin.  

\section{A strategy for deriving bit commitment from coin tossing} 

Abstractly, the basic idea is this.  $A$ and $B$ identify some suitable
graded class $C = \oplus_{n \geq 0} C_n $ of mathematical objects
with the property that there is some increasing function $f(n)$ such 
that the members of $C_n$ can be identified by $f(n)$ bits.  
They also identify a class $D = \oplus_{ n \geq 0} D_n $ of mathematical 
objects, with a relation $ \rightarrow $ defining a subset of 
$D \times C$: we say $d \in D$ is {\it associated to} $c \in C$ if 
$d \rightarrow c$.  
Before implementing the protocol, they will agree on security
parameters $m$ and $n$, and on bit string 
representations for the members of $C_m$ and $D_n$.   
They then carry out $2 f(m)$ secure coin tossings, which they use to generate 
two randomly chosen elements $c_0$ and $c_1$ of $C_m$.  

To commit to a bit $a$, $A$ should then randomly choose  
a member $d$ of $D_n$ such that $d \rightarrow c_a$, and sends
the bit string representation of $d$ to $B$.  To unveil the bit
$a$, $A$ sends $B$ a description of $d$ and a proof that 
$d \rightarrow c_a$.  

Several things are required
for this to define a computationally secure bit commitment
protocol.  

First, it must be hard for $A$ to identify any elements
$d$ such that $d \rightarrow c_0$ and $d \rightarrow c_1$, and
such that she has any significant chance of being in a 
position to prove whichever of these 
results she chooses at the time of unveiling.  This must be
true whether or not her choice of $d$ is in fact random.  
One way of ensuring this would be to ensure that the 
probability of her, at commitment, being able to choose 
any $d$ associated with both $c_a$ is very low or zero.   
Another would be to ensure that, whatever strategy she uses to
choose $d$ initially, 
and whatever strategy she follows during the protocol, her chances --- call
them $p_0$ and $p_1$ --- of 
generating proofs that $d \rightarrow c_0$ and $d \rightarrow c_1$
during the protocol, obey $p_a \leq P_a$, where the numbers $P_a$ 
are fixed by her initial strategy in choosing $d$, and where they obey 
$P_0 + P_1 \leq 1 + \epsilon$, for some suitably small value of
the security parameter $\epsilon$.    

Second, it must be hard for $B$, given a randomly 
chosen $d \rightarrow c_a$, to obtain significant 
information during the protocol about whether $d$ is likelier to be 
associated with $c_a$ or $c_{\bar{a}}$.  

Finally, for the protocol to be practical, it must be easy for $A$
to choose random members of $D_n$ that are associated to a 
randomly chosen $c_a$, by a method which easily generates
a proof of the association.  Also, the proof itself must be
easy to communicate.   

\section{Bit commitment from coin tossing: possible implementations} 

\subsection{Subgraph isomorphism}

One possible implementation is given by creating
random graphs on which $A$ can define instances of the 
graph subisomorphism problem.  
Take $C_m$ and $D_n$ to be the sets of graphs with $m$ and $n$ vertices, 
respectively, with the relation $ d \rightarrow c$ if and only if 
$d$ is a subgraph of $c$.  

With these definitions, and having agreed security parameters $m$ and $n$
with $m>n$, $A$ and $B$ 
generate two random graphs in $C_m$ by carrying out $ m (m-1)$
coin tosses, one for each pair of vertices, and including 
an edge $(i,j)$ if and only if the corresponding coin toss
has result $1$.  $A$ can choose a random subgraph $d \in D_n$
of either graph $c_a \in C_m$ by choosing a random size $n$ 
subset $I = \{i_1 , \ldots , i_n \}$ of the vertices $\{ 1 , \ldots , m \}$ 
of $c_a$.  ($I$ is a random ordered set, i.e. the ordering of the $i_j$ is
randomly chosen; in particular, thus, it is generally not numerical.)    
To send $B$ a description of $d$, she sends the list 
$\{ (k,l ) : ( i_k , i_l ) {\rm~an~edge~of~} c_a \}$.  
To prove to $B$ that $d \rightarrow c_a$, she simply lists the 
ordered subset, allowing $B$ to check the above procedure has 
been followed.   

\subsection{Subset sum}

Another implementation is given by generating random sets of 
positive integers on which $A$ can define instances of subset sum 
problems, defined on sets of density close to $1$.   
Take $C_m$ to be the class of sets of the form
$c = \{ c_1 , \ldots , c_m \}$, where the $c_i$ are 
positive binary integers of length $\leq m$.  Let $D_n$ be the 
set of positive integers less than $n 2^n$.  
Define the relation $ d \rightarrow c$ 
to hold if and only if there is a set of integers $x_i \in \{ 0, 1 \}$
such that $d = \sum_i x_i c_i $.  

With these definitions, having agreed a fixed $n$, $A$ and $B$ can
use $2 m^2$ coin tosses to generate two independent random size $m$ 
sets, $c_0 = \{ a^0_i \}$ and $c_1 = \{ a^1_i \}$, 
using each coin toss to define a specified
bit of a specified set element.  
$A$ can then choose a random set of bits $ x_i \in \{ 0,1 \}$ and 
commit the bit $a$ to $B$ by sending the sum $ d = \sum_i c^a_i x_i$. 
To prove to $B$ that $d \rightarrow c_a$, she simply sends an ordered 
list of the $x_i$.   

\subsection{Remark} 

Other possible implementations could be based on matrix
representability or other problems that are defined by probability
distributions generated by finite strings of coin tosses
and are known to be average case
intractable.\cite{gurevich,vr} 

\section{Security discussion} 

Are these protocols computationally secure against quantum computers?  
Are they computationally secure even against classical 
computers?  These are hard questions.  Proving affirmative
answers would mean proving that $(BQ)P \neq NP$.   
And even assuming that $P \neq NP$ and
$BQP \neq NP$, conjectures which are widely believed, would 
not imply classical or quantum computational security.  
I give here only a short illustrative list of security worries,
folllowed by some reasons for thinking that the protocols might,
nonetheless, be hard for quantum computers to break.

\subsection{Security worries} 

Consider first security against classical computers.  
Recall that both the subgraph isomorphism and
subset sum problems are NP-complete.  (See for example Ref. 
\cite{gareyjohnson}.)   
Let us assume that, as is widely believed, P$\neq$NP.  
If so, and if $B$'s task were to decide whether a graph $d$ was isomorphic
to a subgraph of a graph $c$, or whether an integer $d$ could be written
as a binary sum $\sum_i x_i c_i$ of knapsack elements, then it 
would be impossible to find an algorithm that solved all such
problems in a time polynomial in the problem parameters.   

$B$'s task is slightly different, though.  He has to decide whether
a graph $d$ is isomorphic to a subgraph of graphs $c_0$ or $c_1$, or
whether an integer $d$ is a subset sum from the set $c_0$ or $c_1$, 
knowing that one or the other is the case.   
It seems unlikely that this decision problem is substantially easier than
the subgraph isomorphism or subset sum problems in worst case, since
it is hard to see how to address the first except by 
trying to solve the second for each of $c_0$ and $c_1$.  
But I know no theorem showing that even these slightly modified
problems are still NP-complete.   

An NP-completeness result would anyway not suffice.  
$B$ needs to solve average case, not worst case, instances.  
So the case for security against $B$
relies on the belief that subset sum, subgraph isomorphism, 
or whatever randomly generated problem is chosen, is average
case hard.  

Provably average case complete problems, which could be used
in the protocols above, are known\cite{gurevich,vr}. 
It is also believed that subset sum is average case hard, for appropriate
parameter choices\cite{in}.    
However --- another potential concern --- these average case results and 
conjectures apply to problem instances chosen from  
different probability distributions from ours. 
Normally, when the average case of a decision problem ---
in our notation: is $d \rightarrow c$? --- 
is considered, one assumes that $c$ and $d$ are independently
randomly generated, with suitable probability distributions.  
Here, we have three structures, $c_0$, $c_1$ and $d$, and while
$c_0$ and $c_1$ are independently randomly generated in a standard
way, $d$ is not.  To ensure that $d \rightarrow c_0$ or $d \rightarrow c_1$,
and that $A$ knows which, we required that $A$ use some random
algorithm which takes the description of $c_a$ --- for her choice
of $a$ --- and constructs a $d$ such that $d \rightarrow c_a$.  
We need it to be hard for $B$ to solve the decision problem, for a 
random instance, {\it even if} he knows the random algorithm 
which $A$ uses to define $d$.  (If security were to rely on $A$ keeping
the algorithm secret, and not just its random input, we would not 
have a completely defined protocol.  Any published
rules which told $A$ exactly how to implement the protocol would
allow $B$ to break it.) 

Obviously, the concerns listed above are still more of a worry when
considering quantum attacks, since quantum computers are for some
purposes more powerful than classical computers, and since quantum
complexity is less well understood than classical complexity.  

\subsection{Why might one nonetheless hope the protocols are secure?} 

The case for security, such as it is, begins from the widely shared
belief that there is no quantum algorithm capable of solving the
general case of an NP-complete problem in polynomial time.   
A commonly cited (e.g. Ref. \cite{nielsenchuang}) 
reason for this belief is that NP-complete problems, 
such as subgraph isomorphism or subset sum, are effectively as hard
as searching for a particular entry in a database whose size grows
exponentially in the length of the problem description.  
The reasoning here is that, if one wants to decide, for instance,
whether a size $n$  
graph $H$ is a subgraph of a size $m$ graph $G$, there may be 
no algorithm substantially better than searching through all the 
$ \frac{m!}{(m-n)! n!} $ subgraphs of $G$ and seeing whether $H$ matches
any of them --- the argument being that algorithms which are 
substantially more efficient than brute force need
some mathematical structure to work with, and this type of problem 
just has too little structure to allow such algorithms.  

This intuition could, of course, be wrong.  If it were right, though, 
it would mean that efficient quantum algorithms for NP-complete problems
could, indeed, be excluded, since we know that the quantum algorithms
can give only a square-root speed-up for database 
search.\cite{bbbv}     

Suppose the intuition is right.  It is tempting to take it
somewhat further.  One might speculate that B's problem --- 
searching for a randomly chosen subgraph of one of two random graphs --- is 
also not substantially
easier than a database search, when $m$ and $n$ are suitably related
and $m$ is large.   Similarly, one might speculate that the problem
facing a dishonest A --- finding a common subgraph of two random graphs ---
is not substantially easier than finding collisions of a random
two-to-one function, a generalised search problem which is also
suspected (though not proven) to be hard for quantum computers.   
If so --- if, in the end, the lack of mathematical structure in each
of these problems allows them to resist quantum attack --- then the 
protocols would indeed be secure.   

\section{Conclusions}  

The ideas above suggest a possible way forward in developing
``quantum-immune'' protocols for general mistrustful cryptographic
tasks.  More immediately, they add to the motivation for extending the
folk wisdom about --- or rigorous bounds on --- the power of quantum
computing.  Can we find good reasons for extending the intuition that
quantum computers cannot break general case NP-complete problems to
average case instances?  Can we extend those intuitions further to
collision-type problems such as identifying a common subgraph of
random graphs?  Or, conversely, and against expectation, could there
be reasons to believe that quantum attacks may indeed be effective in
these last two cases?

\section{Acknowledgements} 

This work was partially supported by the European collaborations
EQUIP and PROSECCO.

\end{multicols} 
\end{document}